\documentclass[aps,prl,preprint,showpacs,groupedaddress]{revtex4}
\usepackage {amssymb}
\usepackage {amsmath}
\usepackage {graphicx}
\usepackage {longtable}

\begin{document}
\title{Effect of charge ordering on superconductivity in high-temperature
superconductors}

\author{Fedor V.Prigara}
\affiliation{Institute of Physics and Technology,
Russian Academy of Sciences,\\
21 Universitetskaya, Yaroslavl 150007, Russia}
\email{fprigara@recnti.uniyar.ac.ru}

\date{\today}

\begin{abstract}

It is shown that charge ordering in layered crystalline metals
causes an increase in the normal state energy of the conducting
layers. If this increase in the energy exceeds the difference
between the superconducting state energy and the normal state
energy of the metal, then the superconducting transition occurs. A
relation between the charge gap in the superconducting phase and
the bandgap width in the undoped insulating phase of
high-temperature superconductors is obtained.

\end{abstract}


\maketitle

Recently, it was shown [1] that charge ordering (`charge-density-wave') and
superconductivity coexist in a layered pnictide superconductor $SrPt_{2}
As_{2} $ related to the high-temperature iron pnictide superconductors, such
as $Ba\left( {Fe_{1 - x} Co_{x}}  \right)_{2} As_{2} $. Earlier, it was
shown [2] that the charge gap is present in all high-temperature
superconductors in which the maximum superconducting transition temperature
$T_{c} $ exceeds the transition temperature $T_{f} \approx \theta _{D}
/\alpha $ of a low-temperature ferroelastic transition (here $\theta _{D} $
is the Debye temperature and $\alpha = 18$ is a constant). Here we consider
an effect of charge ordering in layered crystalline metals on
superconductivity with account for a ferroelastic distortion in the
conducting layers induced by charge ordering in adjacent charge-ordered
layers. Charge ordering is necessary for improper high-temperature
superconductivity (with the transition temperature $T_{c} > T_{f} $), but it
is not sufficient. All improper high-temperature superconductors (cuprates,
iron pnictides, barium bismuthate doped with \textit{K}, and possible
high-temperature superconductors $Ca_{2} RuO_{4 + \delta}  $and $Sr_{2}
RuO_{4 + \delta}  $) are superconducting semiconductors [3]. Here we obtain
a relation between the charge gap in the superconducting phase and the
bandgap width in the undoped insulating phase of improper high-temperature
superconductors. We obtain also the value of the optimal doping level in
hole-doped high-temperature superconductors and in electron-doped iron
pnictide superconductors.

Charge ordering implies a local metal-insulator transition in some layers of
a layered crystalline metal with an opening of the charge gap in the
corresponding energy band. The energy bands associated with the conducting
layers remain metallic. Charge ordering causes a ferroelastic lattice
distortion in chrge-ordered layers corresponding to a relative expansion of
the lattice at zero temperature (0 K) [3]. This ferroelastic distortion
leads to a periodic lattice modulation due to a misfit between the lattice
parameters of metallic layers and charge-ordered layers, respectively, below
the charge ordering temperature $T_{s} $.

For a sinusoidal lattice modulation, the displacement \textit{u} of atoms
from their mean locations is given by the formula

\begin{equation}
\label{eq1}
u = u_{0} sin\left( {\frac{{2\pi} }{{P}}x} \right),
\end{equation}

\noindent
where \textit{P} is the period of the lattice modulation along the x-axis.
The length \textit{l} of a period of the sinusoid is given by the formula

\begin{equation}
\label{eq2}
l = P\left( {1 + \frac{{\pi ^{2}u_{0}^{2}} }{{P^{2}}}} \right).
\end{equation}

Thus, the amplitude $\delta = \left( {l - P} \right)/P$ of a ferroelastic
distortion along the x-axis is related to the amplitude $u_{0} /P$ of the
lattice modulation as follows

\begin{equation}
\label{eq3}
\delta = \frac{{\pi ^{2}u_{0}^{2}} }{{P^{2}}}.
\end{equation}

The amplitude $u_{0} $ of the atomic displacement has an order of the
Born-Mayer parameter $\rho = 0.036nm$ [4]. For example, in $Bi_{2} Sr_{2}
CaCu_{2} O_{8 + \delta}  $, the amplitude of the atomic displacement is
$u_{0} = 0.04nm$ [5]. Since the period of the lattice modulation in $Bi_{2}
Sr_{2} CaCu_{2} O_{8 + \delta}  $ along the a-axis (atoms are displaced from
their mean locations in the c-axis direction) is $P = 2.6nm$, the relation
(\ref{eq3}) gives the amplitude of a ferroelastic a-axis distortion at a level of
$\delta = 2.4 \times 10^{ - 3}$.

This value is close to $\delta _{0} = a_{0} /d_{c} $, where $a_{0} = 0.45nm$
has an order of the lattice parameter, and $d_{c} = 180nm$ is the size of a
crystalline domain [2]. The amplitude of a ferroelastic distortion in
$Bi_{2} Sr_{2} CaCu_{2} O_{8 + \delta}  $ has an order of the amplitude of
ferroelastic fluctuations in the superconducting phase just below the
transition temperature $T_{c} $ [2].

In a layered tetragonal compound $LaAgSb_{2} $ which exhibits a
charge-ordering (`charge-density-wave') transition at $T_{s} = 210K$ [6],
the period \textit{P} of the lattice modulation along the a-axis is $P =
16.6nm$. If the amplitude $u_{0} $ of the atomic displacement has an order
of the Born-Mayer parameter $\rho = 0.036nm$, then the equation (\ref{eq3}) gives
the amplitude of the a-axis ferroelastic distortion at a level of $\delta
\cong 0.5 \times 10^{ - 4}$ (or $\delta \cong \frac{{1}}{{4}}\rho /d_{c} $).
Such is an order of the amplitude of a low-temperature ferroelastic
distortion in metals and insulators [3, 7, 8].

A pnictide superconductor $SrPt_{2} As_{2} $ exhibits a charge-ordering
(`charge-density-wave') transition at $T_{s} = 470K$ [1]. The amplitude
$\delta $ of the orthorhombic distortion is

\begin{equation}
\label{eq4}
\delta = 2\frac{{b - a}}{{b + a}} \cong 0.011.
\end{equation}

The period \textit{P} of the lattice modulation along the a-axis is $P =
0.72nm$. The equation (\ref{eq3}) gives an estimation of the amplitude $u_{0} $ of
the atomic displacement in the form

\begin{equation}
\label{eq5}
u_{0} \cong \frac{{P}}{{\pi} }\delta ^{1/2} \cong 0.024nm,
\end{equation}

\noindent
or $u_{0} \cong \frac{{2}}{{3}}\rho $, where $\rho $ is the Born-Mayer
parameter.

The energy pseudogap associated with a multi-band structure of the
electronic spectrum (an analogue of the charge gap in layered compounds) can
exist even in elemental metals. In \textit{3d} transition metals, the
\textit{d}-band consists of two subbands [9], corresponding to $e_{g} $ and
$t_{2g} $ states in the splitting by the crystal electric field picture [4].
The magnitude of the energy pseudogap in \textit{3d} transition metals is
about $0.5eV$. Magnetic ordering (in Cr, Mn, Fe, Co, and Ni) is associated
with the filling of the upper subband. The 4-state subband ($e_{g} $ states)
stabilizes a bcc crystal structure, and the 6-state subband ($t_{2g} $
states) stabilizes fcc and hcp crystal structures.

Absence of the superconducting transition down to zero temperature
(\textit{T=0 K}) in a metal means that the energy $E_{s} $ of the
superconducting state is higher than the energy $E_{n} $ of the normal state
due to the compensation of a low-temperature ferroelastic distortion in the
superconducting phase [3]. (A low-temperature ferroelastic distortion causes
a decrease in the value of the energy $E_{n} $ of the normal state in
metals).

The energy \textit{E} of a layered crystalline metal is

\begin{equation}
\label{eq6}
E = E_{1} + E_{2} + E_{12} ,
\end{equation}

\noindent
where $E_{1} $ is the energy of the charge-ordered layers, $E_{2} $ is the
energy of the conducting layers, and $E_{12} $ is the energy of interaction
between these layers.

Charge ordering (a local metal-insulator transition in some layers) causes a
decrease in the energy $E_{1} $ of the charge-ordered layers. However, the
energy $E_{2} $ of the conducting layers increases due to a ferroelastic
distortion in these layers induced by charge ordering in adjacent layers
(the epitaxial strain effect), since a low-temperature ferroelastic
distortion has opposite signs in metals and insulators [3]. If an increase
$\Delta E_{2} $ in the energy of the conducting layers exceeds the
difference $E_{2s} - E_{2n} $ of the energies of the superconducting state
and the normal state, respectively,

\begin{equation}
\label{eq7}
\Delta E_{2} > E_{2s} - E_{2n} ,
\end{equation}

\noindent
then the superconducting transition occurs.

A total energy \textit{E} of a layered metal decreases (on cooling) both at
the charge-ordering transition and at the superconducting transition.

A magnitude $\Delta _{ch} \left( {0} \right)$ of the charge gap at zero
temperature ($0K$) is related to the charge-ordering transition temperature
$T_{s} $ as follows [3]

\begin{equation}
\label{eq8}
\Delta _{ch} \left( {0} \right) = \alpha k_{B} T_{s} ,
\end{equation}

\noindent
where $k_{B} $ is the Boltzmann constant, and $\alpha = 18$.

A ferroelastic distortion associated with the metal-insulator transition
(and charge ordering) corresponds to a relative expansion of the lattice at
zero temperature ($0K$). In ZnO with a wurtzite crystal structure, the
bandgap width $E_{g} \left( {0} \right)$ at zero temperature is $E_{g}
\left( {0} \right) = 3.44eV$ [10], and the relation analogues to the
equation (\ref{eq8}) gives the metal-insulator transition temperature $T_{MI} $ in
the form $T_{MI} = E_{g} \left( {0} \right)/\alpha k_{B} = 2216K$. A
ferroelastic distortion associated with this high-temperature
metal-insulator transition in ZnO is a c-axis distortion. The \textit{c/a}
ratio increases with decreasing temperature ($c/a = 1.59$ at $T = 1500K$,
and $c/a = 1.60$ at $T = 0K$). A ferroelastic distortion in the insulating
phase produces a negative thermal expansion at low temperatures. A linear
thermal expansion coefficient along the a-axis $\alpha _{a} $ is negative
below $100K$, and a linear thermal expansion coefficient along the c-axis
$\alpha _{c} $ is negative below $120K$ [10].

A relative expansion of the lattice at zero temperature ($0K$) in insulators
causes a decrease in the volume of the Brillouin zone. The valence band
contracts, and the energy of a crystal decreases. Similar is the effect of a
relative expansion of the lattice at zero temperature associated with the
superconducting transition, due to the opening of the superconducting energy
gap.

In metals, a relative contraction of the lattice at zero temperature causes
an increase in the volume of the Brillouin zone, so that the conduction band
expands on the energy scale, and the energy of the bottom of the conduction
band decreases. As a result, the energy of a metal decreases due to a
low-temperature ferroelastic distortion. A ferroelastic distortion
associated with the superconducting transition compensates a low-temperature
ferroelastic distortion in a meta [3], so that the energy $E_{s} $ of the
superconducting state can be both higher or lower than the energy $E_{n} $
of the normal state.

A ferroelastic distortion associated with ferromagnetic or ferrimagnetic
ordering normally corresponds to a relative expansion of the lattice at zero
temperature, for example, in the nickel chromite spinel $NiCr_{2} O_{4} $
[11]. Antiferromagnetic ordering in insulators is normally associated with a
relative contraction of the lattice at zero temperature, for example, in
manganese fluoride $MnF_{2} $ [7].

In optimally doped improper high-temperature superconductors, the
superconducting transition coincides with the charge-ordering transition, so
that a magnitude $\Delta _{ch} \left( {0} \right)$ of the charge gap at zero
temperature is determined by the equation

\begin{equation}
\label{eq9}
\Delta _{ch} \left( {0} \right) = \alpha k_{B} T_{c} ,
\end{equation}

\noindent
where $T_{c} $ is the maximum superconducting transition temperature.

In narrow bandgap semiconductors, there is a relation between the energy
$E_{i} \left( {0} \right)$ of an elementary insulating excitation, which
determines the metal-insulator transition temperature $T_{MI} $, and the
bandgap width $E_{g} \left( {0} \right)$ at zero temperature of the form [3]

\begin{equation}
\label{eq10}
E_{i} \left( {0} \right) = zE_{g} \left( {0} \right),
\end{equation}

\noindent
where \textit{z} is the coordination number.

There is a similar relation between the bandgap width $E_{g} \left( {0}
\right)$ in the undoped insulating phase and the charge gap $\Delta _{ch}
\left( {0} \right)$ at zero temperature in the superconducting phase of
improper high-temperature superconductors,

\begin{equation}
\label{eq11}
E_{g} \left( {0} \right) = z\Delta _{ch} \left( {0} \right).
\end{equation}

The bandgap width $E_{g} \left( {0} \right)$ in the insulating phase (at
zero temperature) is determined by the metal-insulator transition
temperature $T_{MI} $, as given by the equation analogues to the equation
(\ref{eq8}),

\begin{equation}
\label{eq12}
E_{g} \left( {0} \right) = \alpha k_{B} T_{MI} .
\end{equation}

For example, in $YBa_{2} Cu_{3} O_{y} $, the metal-insulator transition
temperature is $T_{MI} = 410K$ at $y = 6.23$ [12], so that the bandgap width
in the undoped insulating phase is $E_{g} \left( {0} \right) = 0.635eV$. The
maximum superconducting transition temperature is $T_{c} = 92K$ at the
optimal doping level of $p_{m} \cong 0.20$ (\textit{p} is the number of
holes per Cu atom), so that a magnitude of the charge gap in the
superconducting phase is $\Delta _{ch} \left( {0} \right) = 0.14eV$.
According to the equation (\ref{eq11}), the coordination number is $z = 5$.

In $La_{2} CuO_{4 + \delta}  $, the metal-insulator transition temperature
is $T_{MI} = 240K$ at $\delta = 0$, so that the bandgap width in the undoped
insulating phase is $E_{g} \left( {0} \right) = 0.37eV$. The maximum
superconducting transition temperature is $T_{c} = 45K$ at the optimal
doping level of $p_{m} \cong 0.20$ ($\delta = 0.093$) [13], so that a
magnitude of the charge gap in the superconducting phase is $\Delta _{ch}
\left( {0} \right) = 0.07eV$. The coordination number is $z = 5$.

In a hole-doped iron pnictide superconductor $Ba_{1 - x} K_{x} Fe_{2} As_{2}
$, an extrapolated value of the metal-insulator transition temperature at $x
= 0$ is $T_{MI} \cong 150K$ [3], so that $E_{g} \left( {0} \right) =
0.23eV$. The maximum superconducting transition temperature is $T_{c} = 38K$
at the optimal doping level of $p_{m} = 0.25$ [14], so that a magnitude of
the charge gap in the superconducting phase is $\Delta _{ch} \left( {0}
\right) = 0.06eV$. The coordination number, according to the equation (\ref{eq11}),
is $z = 4$.

In all these cases, a value $p_{m} $ of the optimal doping level in a
hole-doped improper high-temperature superconductor is related to the
coordination number \textit{z} as follows

\begin{equation}
\label{eq13}
p_{m} = 1/z.
\end{equation}

According to the equation (\ref{eq11}), this relation gives

\begin{equation}
\label{eq14}
\Delta _{ch} \left( {0} \right) = p_{m} E_{g} \left( {0} \right).
\end{equation}

In $Ba\left( {Fe_{1 - x} Co_{x}}  \right)_{2} As_{2} $, the magnitude
$\Delta _{ch} \left( {0} \right)$ of the charge gap was directly measured by
means of scanning tunneling spectroscopy [15]. Experimental values of the
charge gap are ranging from $24meV$ to $40meV$. Since the maximum
superconducting transition temperature in this electron-doped iron pnictide
high-temperature superconductor is $T_{c} = 23K$ [16], the equation (\ref{eq9})
gives $\Delta _{ch} \left( {0} \right) = 36meV$. The magnitude $\Delta
\left( {0} \right)$ of the superconducting gap is determined by the equation
[17]

\begin{equation}
\label{eq15}
2\Delta \left( {0} \right) = \alpha _{P} \alpha k_{B} T_{c} ,
\end{equation}

\noindent
where $\alpha _{P} = 3/8$ for layered high-temperature superconductors [2].
The equation (\ref{eq15}) gives for $Ba\left( {Fe_{1 - x} Co_{x}}  \right)_{2}
As_{2}  \quad 2\Delta \left( {0} \right) = 13meV$. This value agrees with
experimental values of the superconducting gap [15].

The coordination number \textit{z} for the transition metal atom in this
compound is $z = 3$ [1]. Since the bandgap width $E_{g} \left( {0}
\right)$in the undoped insulating phase is $E_{g} \left( {0} \right) =
0.23eV$ (see above), the charge gap $\Delta _{ch} \left( {0} \right)$in the
superconducting phase is related to $E_{g} \left( {0} \right)$ in
electron-doped iron pnictide superconductors as follows

\begin{equation}
\label{eq16}
E_{g} \left( {0} \right) = 2z\Delta _{ch} \left( {0} \right).
\end{equation}

Due to a two-phase composition of the $Ba\left( {Fe_{1 - x} Co_{x}}
\right)_{2} As_{2} $ system even in the nearly optimally doped region
[3,18], it is difficult to determine the optimal doping level in
electron-doped iron pnictide superconductors. It seems that the mean optimal
doping level $\bar {n}_{m} $ (\textit{n} is the number of electrons per Fe
atom) in these superconductors is given by the formula

\begin{equation}
\label{eq17}
\bar {n}_{m} = 1/\left( {4z} \right).
\end{equation}

In $SrPt_{2} As_{2} $, only one of the two \textit{PtAs} layers within a
unit cell is conducting and another \textit{PtAs} layer exhibits charge
ordering [1]. If we assume that in the $Ba\left( {Fe_{1 - x} Co_{x}}
\right)_{2} As_{2} $ system only one of the two \textit{FeAs} layers within
a unit cell is conducting and another \textit{FeAs} layer is charge-ordered,
the a real optimal doping level $n_{m} $ is

\begin{equation}
\label{eq18}
n_{m} = 2\bar {n}_{m} = 1/\left( {2z} \right).
\end{equation}

In view of the equation (\ref{eq16}), the last relation gives

\begin{equation}
\label{eq19}
\Delta _{ch} \left( {0} \right) = n_{m} E_{g} \left( {0} \right).
\end{equation}

The equation (\ref{eq19}) is analogues to the equation (\ref{eq14}) for hole-doped
high-temperature superconductors.

In a hole-doped iron pnictide superconductor $Ba_{1 - x} K_{x} Fe_{2} As_{2}
$, the coordination number $z = 4$ can be attributed to the As atoms [1].

A pairing mechanism for high-temperature superconductivity as well as for
low-temperature superconductivity is presumably the interaction of electrons
with ferroelastic fluctuations. Ferroelastic fluctuations are always present
in the superconducting phase of both low-temperature and high-temperature
superconductors [2]. Antiferromagnetic fluctuations are also present in the
superconducting phase [17], however, only in paramagnetic metals [2]. An
improper high-temperature superconductor $Ba_{1 - x} K_{x} BiO_{3} $ with
$T_{c} = 30K$ [19] does not exhibit antiferromagnetic ordering in the
undoped insulating phase.

To summerize, we show that charge ordering (a local metal-insulator
transition) in some layers of layered high-temperature superconductors
induces the superconducting transition in adjacent conducting layers. The
lattice modulation in layered high-temperature superconductors is produced
by a misfit in the lattice parameters between the charge-ordered layers and
the conducting layers.. There is a relation between the charge gap in the
superconducting phase and the bandgap width in the undoped insulating phase
of high-temperature superconductors. There is also a relation between the
value of the optimal doping level and coordination number. In electron-doped
iron pnictide superconductors, only one of the two \textit{FeAs} layers in a
unit cell is conducting and another \textit{FeAs} layer is charge-ordered.

\begin{center}
---------------------------------------------------------------
\end{center}

[1] K.Kudo, Y.Nishikubo, and M.Nohara, J.Phys. Soc. Jpn. (submitted),
arXiv:1010.3950 (2010).

[2] F.V.Prigara, arXiv:1001.4152 (2010).

[3] F.V.Prigara, arXiv:1001.3061 (2010).

[4] V.S.Urusov, \textit{Theoretical Crystal Chemistry} (Moscow University
Press, Moscow, 1987).

[5] J.A.Slezak, J.Lee, M.Wang et al., Proc. Nat. Acad. Sci. USA
\textbf{105}, 3203 (2008).

[6] S.L.Bud'ko, S.A.Law, P.C.Canfield, G.D.Samolyuk, M.S.Torikachvili, and
G.M.Schmiedeshoff, J.Phys.: Condens. Matter \textbf{20}, 115210 (2008).

[7] R.Schleck, Y.Nahas, R.P.S.M.Lobo, J.Varignon, M.B.Lepetit, C.S.Nelson
and R.L.Moreira, Phys. Rev. B \textbf{82}, 054412 (2010).

[8] D.T.Adroja, A.D.Hillier, P.P.Deen et al., Phys. Rev. B \textbf{82},
104405 (2010).

[9] E.M.Sokolovskaya and L.S.Guzei, \textit{Metal Chemistry} (Moscow
University Press, Moscow, 1986).

[10] H.Morkoc and U.Ozgur, \textit{Zinc Oxide: Fundamentals, Materials, and
Device Technology} (Wiley-VCH, Berlin, 2009).

[11] H.Ishibashi and T.Yasumi, J.Magn. Magn. Mater. \textbf{310}, e610
(2007).

[12] Y.Wang and N.P.Ong, Proc. Nat. Acad. Sci. USA \textbf{98}, 11091
(2001).

[13] T.Hirayama, M.Nakagawa, and Y.Oda, Solid State Commun. \textbf{113},
121 (1999).

[14] J.K.Dong, L.Ding, H.Wang, X.F.Wang, T.Wu, X.H.Chen, and S.Y.Li, New
J.Phys. \textbf{10}, 123031 (2008).

[15] H.Zhang, J.Dai, Y.Zhang et al., Phys. Rev. B \textbf{81}, 104520
(2010).

[16] S.L.Bud'ko, N.Ni, and P.C.Canfield, Phys. Rev. B \textbf{79}, 220516(R)
(2009).

[17] F.V.Prigara, arXiv:0708.1230 (2007).

[18] J.-H. Chu, J.G.Analytis, K.De Greve, P.L.McMahon, Z.Islam, Y.Yamamoto,
and I.R.Fisher, arXiv:1002.3364 (2010).

[19] D.G.Hinks, B.Dabrowski, J.D.Jorgensen, A.W.Mitchell, D.R.Richards,
S.Pei, and D.Shi, Nature \textbf{333}, 836 (1988).

\end{document}